\documentclass[compsoc, conference, a4paper, 10pt, times]{IEEEtran}

\usepackage{cite}
\usepackage{amsmath,amssymb,amsfonts}
\usepackage{algorithmic}
\usepackage{graphicx}
\usepackage{textcomp}
\usepackage{xcolor}
\usepackage{booktabs}
\usepackage[hidelinks]{hyperref}
\usepackage{subfigure}

\begin{document}

\title{An investigation of Quality Issues in Vulnerability Detection Datasets}


\author{\IEEEauthorblockN{1\textsuperscript{st} Yuejun Guo}
\IEEEauthorblockA{\textit{Luxembourg Institute of Science and Technology} \\
Esch-sur-Alzette, Luxembourg \\
yuejun.guo@list.lu}
\and
\IEEEauthorblockN{2\textsuperscript{nd} Seifeddine Bettaieb}
\IEEEauthorblockA{\textit{Luxembourg Institute of Science and Technology} \\
Esch-sur-Alzette, Luxembourg \\
seifeddine.bettaieb@list.lu}
}

\maketitle

\begin{abstract}
Vulnerability detection is a crucial yet challenging task to identify potential weaknesses in software for cyber security. Recently, deep learning (DL) has made great progress in automating the detection process. Due to the complex multi-layer structure and a large number of parameters, a DL model requires massive labeled (vulnerable or secure) source code to gain knowledge to effectively distinguish between vulnerable and secure code. In the literature, many datasets have been created to train DL models for this purpose. However, these datasets suffer from several issues that will lead to low detection accuracy of DL models. In this paper, we define three critical issues (i.e., data imbalance, low vulnerability coverage, biased vulnerability distribution) that can significantly affect the model performance and three secondary issues (i.e., errors in source code, mislabeling, noisy historical data) that also affect the performance but can be addressed through a dedicated pre-processing procedure. In addition, we conduct a study of 14 papers along with 54 datasets for vulnerability detection to confirm these defined issues. Furthermore, we discuss good practices to use existing datasets and to create new ones.
\end{abstract}



\section{Introduction}
Detecting vulnerabilities plays a vital role in protecting software applications from potential security threats. By automating the detection process, organizations can quickly respond and mitigate risks, which is especially important given the increasing number of vulnerabilities in today's digital landscape. According to the Common Vulnerabilities and Exposures (CVE) Details\footnote{\url{https://www.cvedetails.com/browse-by-date.php}}, 25,227 vulnerabilities were been reported in 2022 and 5910 have already emerged within three months in 2023.

Vulnerability detection is a challenging task that requires the analysis of vast amounts of code, often in various programming languages and with different styles and structures. Traditional approaches to vulnerability detection, such as rule-based~\cite{Lee2016rule} and signature-based~\cite{janaka2023android} methods, rely on human experts to pre-define a set of rules or patterns, which can be time-consuming to develop and may not be effective in identifying new or previously unknown vulnerabilities. By contrast, deep learning (DL) has been proven to be a promising alternative given its ability to automatically learn complex patterns and features from a large amount of source code~\cite{devign2019zhou,codexglue2021paper,saikat2022far}. 

Despite the significant progress in DL-based vulnerability detection, the availability of high-quality benchmark datasets remains a major challenge. As highly data-driven, DL models require a large number of data in the training procedure to tune parameters. Several datasets have been created and open-sourced in the literature (see Table~\ref{tab:statis} for more details) to facilitate the study on vulnerability detection. However, the quality of these datasets is often poor, which can lead to inaccurate, biased, or incomplete results of DL models for vulnerability detection. For instance, the quality of datasets can refer to the coverage of vulnerability types (please refer to Section~\ref{subsubsec:coverage} for more details). If the vulnerability of denial of service (DoS) is not involved in the training set, a detection model is unlikely to identify a code sample with this vulnerability as vulnerable since no related knowledge has been gained. 

In this paper, we aim to reveal issues in existing datasets that need to be aware of when creating new datasets and addressed when using existing ones to improve the effectiveness of vulnerability detection models. Specifically, based on 54 existing datasets from 14 papers about vulnerability detection, we define two types of issues, critical and secondary, based on the degree that an issue can affect the detection performance and the difficulty to be addressed. 
Such critical issues include small sampling size and data imbalance, low coverage of vulnerability types, and bias in vulnerability distribution. Secondary issues, on the other hand, may also affect the model performance but can be addressed through a careful check and pre-processing before feeding into training a DL model. These issues include errors in raw data, mislabeling on source code, and noisy historical data. Finally, based on our findings, we provide actionable suggestions to researchers when using existing datasets as well as creating new ones.

\section{Background and Related Work}
DL-based vulnerability detection has become an increasingly important component of software security as it enables developers and security professionals to identify potential vulnerabilities more quickly and accurately than manual testing. The process involves the training of a DL model that analyzes source code for known patterns that may indicate the presence of a vulnerability, such as buffer overflows, SQL injection, or cross-site scripting~\cite{cve_details_web}. Various DL models have been developed and proved to perform effectively on given test code, such as the graph neural network-based model Devign~\cite{devign2019zhou} and the large-scale pre-trained model CodeBERT~\cite{codexglue2021paper}. 

Nonetheless, the effectiveness of DL-based vulnerability detection models highly depends on the quality of the data that is used for training. If the data is incomplete, inconsistent, or biased, the models may produce unreliable results leading to a false sense of security in addition to wasted resources. Therefore, ensuring the quality of data is essential for the success of vulnerability detection.

The quality issue in existing datasets has attracted the attention of researchers. A recent study~\cite{croft2023data} found that 20-71\% of vulnerabilities are inaccurate in its considered four existing datasets and 17-99\% data is duplicated. Jimenez \emph{et al.}~\cite{Jimenez2019importance} demonstrated that noisy historical data that is labeled secure but actually is undiscovered vulnerabilities can cause the detection accuracy decrease by over 20\%. Garg \emph{et al.}~\cite{garg2022learning} confirmed that taking the nosisy historical data into consideration helps to improve a DL model's performance on vulnerability prediction.

\section{Issues in Existing Datasets For Vulnerability Detection}
We categorize the issues in existing datasets for DL-based vulnerability detection into two types, critical and secondary, based on their impact on training a DL model with SOTA performance and the difficulty to be addressed by data pre-processing. Here, pre-processing refers to common operations to prepare data before training DL models, such as removing comments, removing empty lines, writing data into a unified format (e.g., JSON and PKL files), and splitting data for training, validation, and testing. Additionally, a study on 14 references listed in Table~\ref{tab:statis} is conducted to instantiate these issues. 

\begin{table*}[ht]
\centering
\caption{Overview of collected datasets. \textbf{real}: source code collected from real-world projects. \textbf{synthetic}: artificially generated code. \textbf{mix}: real+synthetic. \textbf{PL}: programming language.}
\label{tab:statis}
\resizebox{\textwidth}{!}{%
\begin{tabular}{ccccccc}
\hline
textbf{Ref.} & \textbf{Year} & \textbf{Type} & \textbf{Granularity} & \textbf{PL} & \textbf{Labeling} & \textbf{Availability} \\ \hline
\cite{Scandariato2014predicting} & 2014 & synthetic & file & Java & known static analysis tool & - \\
\cite{Lin2017discovery} & 2017 & real & function & C & manual & \url{https://github.com/DanielLin1986/function\_representation\_learning} \\
\cite{Choi2017end} & 2017 & synthetic & function & C & manual & \url{https://github.com/mjc92/buffer\_overrun\_memory\_network} \\
\cite{le2018maximal} & 2018 & mix & binary function & C & manual & \url{https://github.com/dascimal-org/MDSeqVAE} \\
\cite{lin2018cross} & 2018 & real & function & C & manual& \url{https://github.com/DanielLin1986/TransferRepresentationLearning} \\
\cite{vuldeepecker2018li} & 2018 & mix & program slice & C/C++ & manual & \url{https://github.com/CGCL-codes/VulDeePecker} \\
\cite{devign2019zhou} & 2019 & real & function & C & manual & \url{https://sites.google.com/view/devign} \\
\cite{miu2019zou} & 2019 & real & program slice & C & manual & \url{https://github.com/muVulDeePecker/muVulDeePecker} \\
\cite{Lin2020deep} & 2020 & real & function & C & manual & \url{https://github.com/Seahymn2019/Function-level-Vulnerability-Dataset} \\
\cite{Liu2020large} & 2020 & real & project & Python & manual & - \\
\cite{sysevr2022li} & 2021 & mix & program slice & C/C++ & manual & \url{https://github.com/SySeVR/SySeVR} \\
\cite{d2a2021zheng} & 2021 & real & trace & C & designed static analysis tool & \url{https://developer.ibm.com/exchanges/data/all/d2a/} \\
\cite{Lin2021multi} & 2021 & mix & function & C & manual & \url{https://github.com/DanielLin1986/RepresentationsLearningFromMulti\_domain} \\
\cite{codexglue2021paper} & 2022 & real & function & C & manual & \url{https://github.com/microsoft/CodeXGLUE/tree/main/Code-Code/Defect-detection/dataset} \\ \hline
\end{tabular}
}
\end{table*}

\subsection{Critical issues}
This category includes inherent issues that can heavily affect a model's performance and cannot be addressed through pre-processing. 

\subsubsection{Small sampling size and data imbalance} 
Since DL models are highly data-driven, feeding sufficient source code into training is crucial when training a DL model for vulnerability detection. However, existing datasets often have a small size. On the other hand, data imbalance has been proven to be a serious problem in the literature~\cite{seperated2016wang,cbl2019cui,focal2020lin}. In vulnerability detection, this imbalance refers to the high ratio of secure code to vulnerable ones.

\begin{figure}[htpb]
    \centering
    \subfigure[Three datasets in~\cite{Lin2017discovery}]
    {
    \label{fig:lin2017_size}
    \includegraphics[scale=0.36]{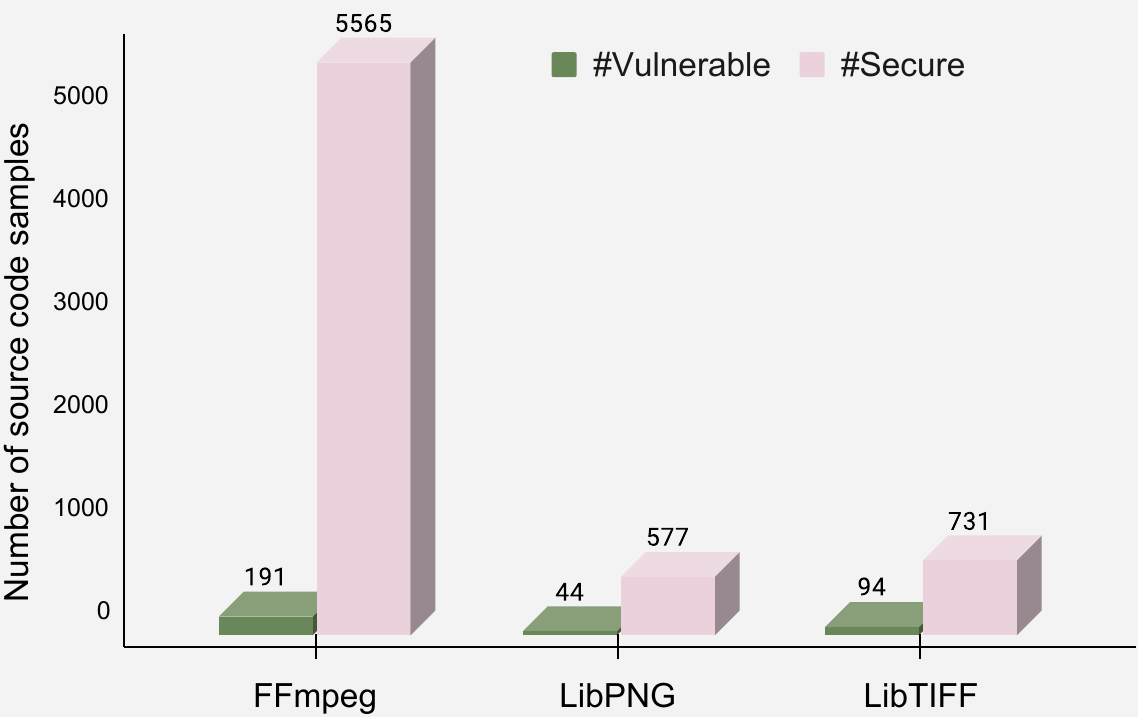}
    }
    \subfigure[Six datasets in~\cite{lin2018cross}]
    {
    \label{fig:lin2018_size}
    \includegraphics[scale=0.36]{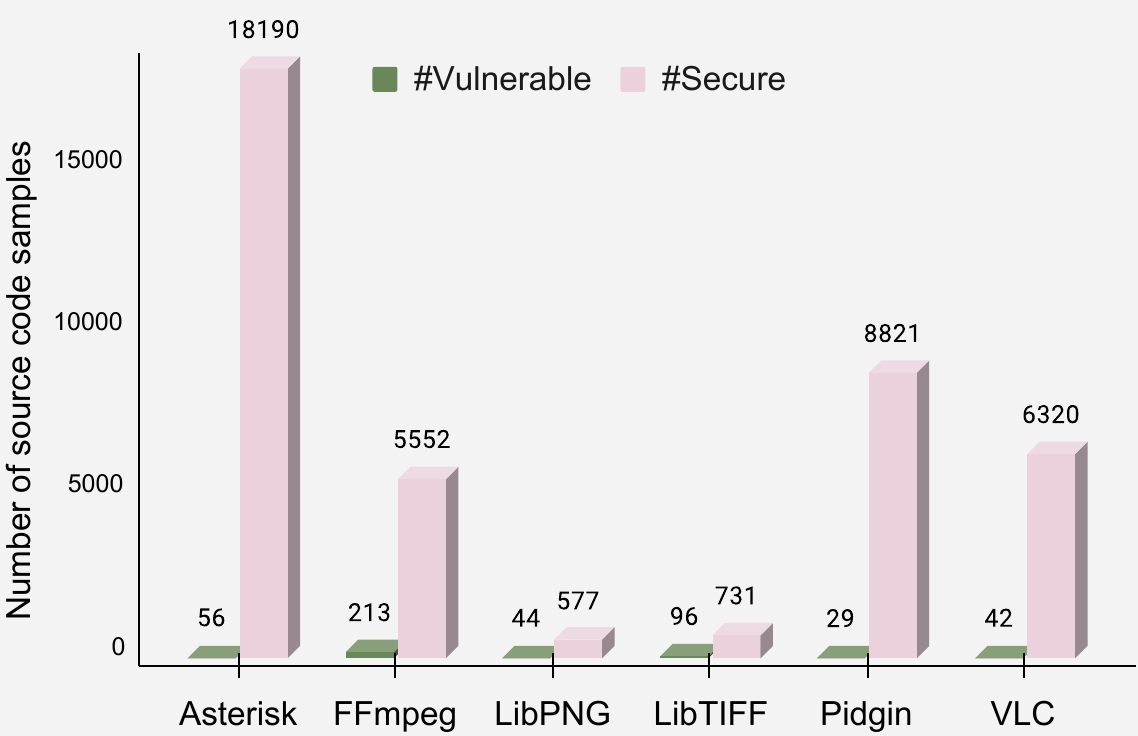}
    }
    \subfigure[Six datasts in~\cite{Lin2021multi}]{
    \includegraphics[scale=0.36]{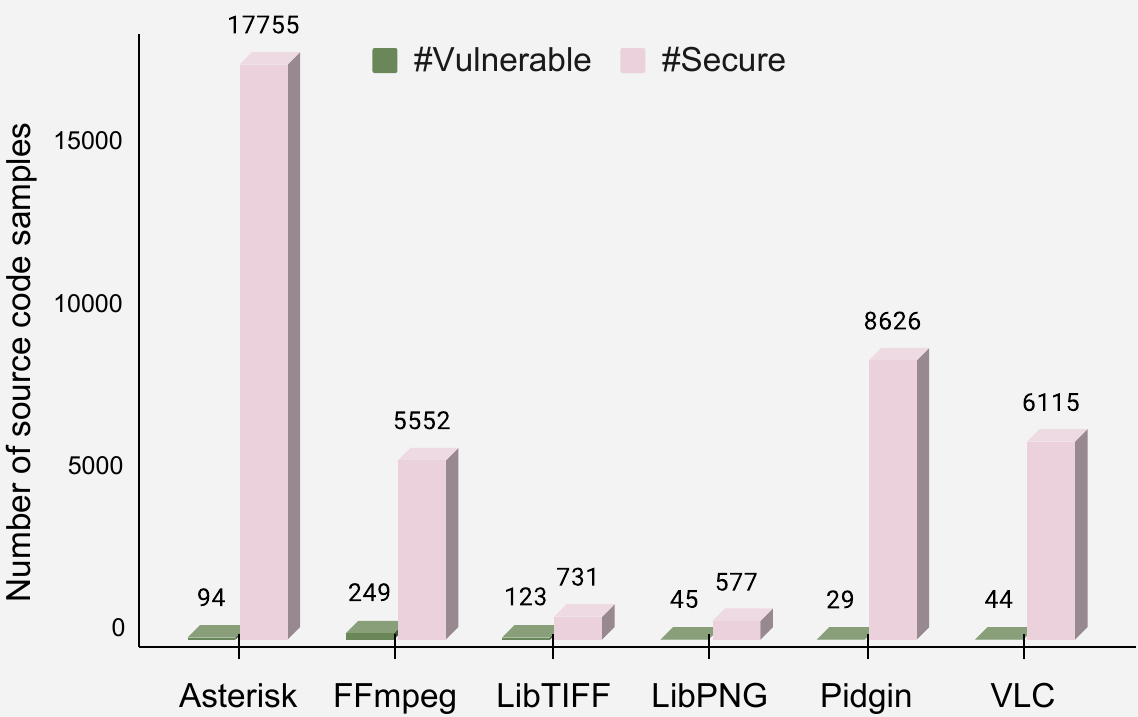}
    }
    \subfigure[Six datasets in~\cite{d2a2021zheng}]{
    \includegraphics[scale=0.36]{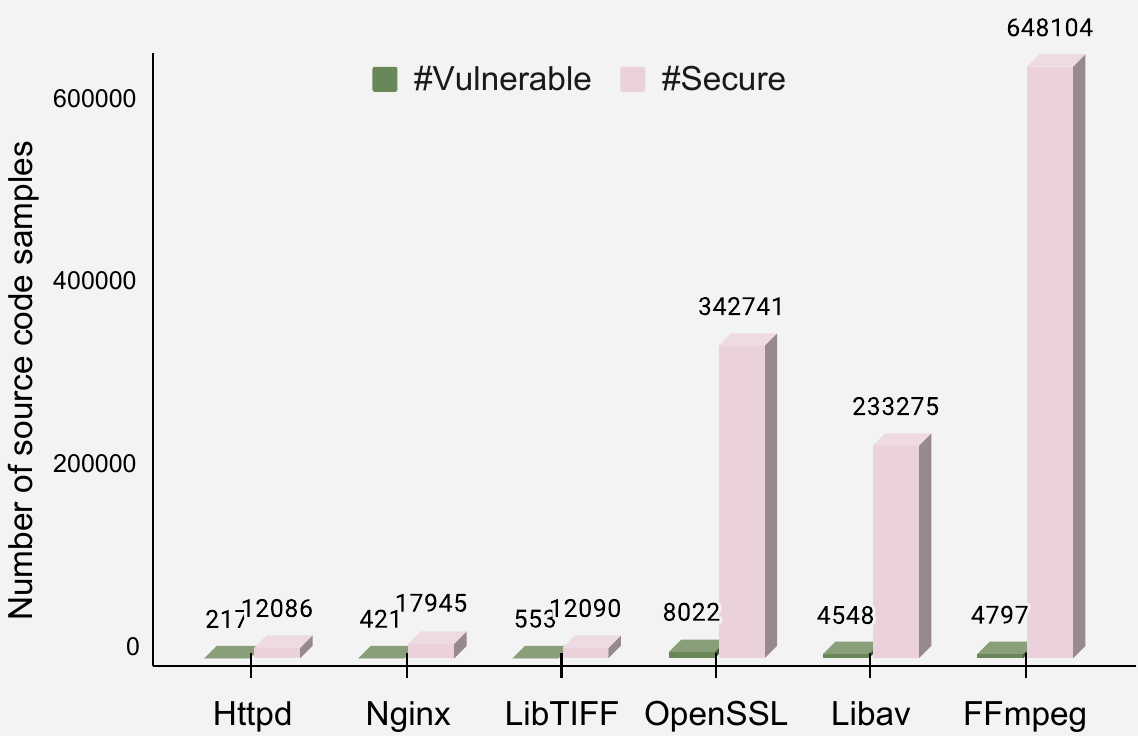}
    }
    \subfigure[\cite{devign2019zhou}: FFmpeg, QEMU; \cite{codexglue2021paper}: Devign; \cite{Lin2020deep}: others]
    {
    \label{fig:other_size}
    \includegraphics[scale=0.36]{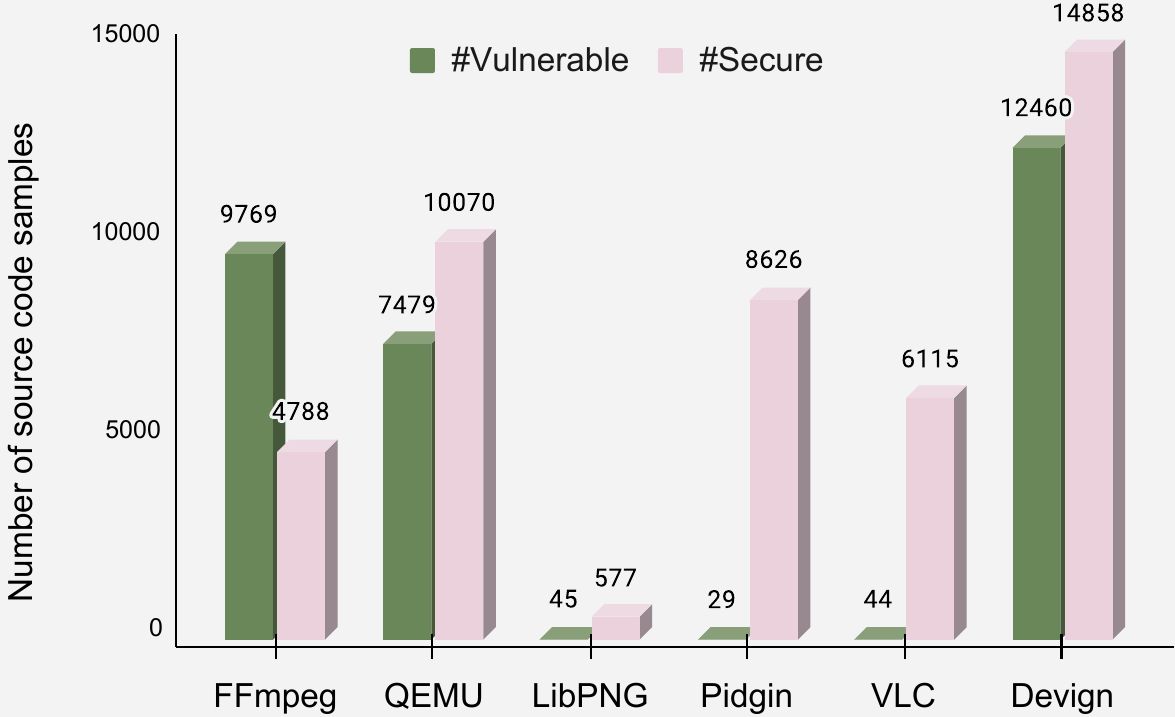}
    }
    \caption{Number of vulnerable and secure codes provided by each reference. $x$-axis shows the project that the data are extracted from.}
    \label{fig:data_size}
\end{figure}

Figure~\ref{fig:data_size} shows the sampling information of 27 datasets provided by seven references. The datasets from two projects, LibPNG and LibTIFF, generally are small-sized. For example, the dataset LibPNG only includes 621 source code samples in total, as shown in Figure~\ref{fig:lin2017_size}. Concerning the sampling bias, except the FFmpeg dataset by~\cite{devign2019zhou} in Figure~\ref{fig:other_size}, all datasets contain more secure code samples than vulnerable ones. Particularly, the imbalance ratio reaches 325 in the Asterisk dataset provided by~\cite{lin2018cross} in Figure~\ref{fig:lin2018_size}. Here, the imbalance ratio is calculated by $\frac{\#Secure}{\#Vulnerable}$.

\subsubsection{Low coverage of vulnerability types}
\label{subsubsec:coverage}
Vulnerability detection models are expected to detect a broad range of vulnerability types, including common ones such as denial of service (DoS) and cross-site scripting (XSS), as well as less common ones such as HTTP response splitting and cross-site request forgery (CSRF) vulnerabilities~\cite{cve_details_web}. However, existing datasets often have limited coverage of vulnerability types, which can lead to false negatives or an overall poor detection. 

Table~\ref{tab:vul_type} lists the vulnerability types included in the 18 datasets provided by four references where the vulnerability of a source code sample is available. In total, 28 vulnerability types are covered by all references, which belong to 10 vulnerability type families~\cite{cve_details_web}, bypass something, XSS, DoS, directory traversal, code execution, gain privileges, overflow, gain information, SQL injection, and file inclusion. Other vulnerability families, such as HTTP response splitting and CSRF, are not covered by any dataset. In addition, note that each vulnerability family includes several vulnerability types, while in the list, most vulnerability types belong to the DoS (ID: 4-13) and the code execution (ID: 5-8, 15-23) vulnerability families. On the other hand, none of the four reverences could cover all listed vulnerability types.

\begin{table}[!htpb]
\centering
\caption{List of vulnerability types included in datasets provided by related references.}
\label{tab:vul_type}
\resizebox{\columnwidth}{!}{%
\begin{tabular}{llcccc}
\hline
\textbf{ID} & \textbf{Vulnerability type} & \textbf{\cite{Lin2017discovery}} & \textbf{\cite{lin2018cross}} & \textbf{\cite{Lin2020deep}} & \textbf{\cite{Lin2021multi}} \\ \hline
1 & Bypass a restriction or similar &  & \checkmark &  & \checkmark \\
2 & Bypass a restriction or similar obtain information &  &  &  & \checkmark \\
3 & Cross-site scripting &  & \checkmark & \checkmark & \checkmark \\
4 & Denial of service & \checkmark & \checkmark & \checkmark & \checkmark \\
5 & Denial of service execute code &  & \checkmark & \checkmark & \checkmark \\
6 & Denial of service execute code memory corruption & \checkmark & \checkmark & \checkmark & \checkmark \\
7 & Denial of service execute code overflow & \checkmark & \checkmark & \checkmark & \checkmark \\
8 & Denial of service execute code overflow memory corruption & \checkmark & \checkmark &  &  \\
9 & Denial of service memory corruption & \checkmark & \checkmark &  & \checkmark \\
10 & Denial of service obtain information & \checkmark & \checkmark &  & \checkmark \\
11 & Denial of service overflow & \checkmark & \checkmark & \checkmark & \checkmark \\
12 & Denial of service overflow memory corruption & \checkmark & \checkmark & \checkmark & \checkmark \\
13 & Denial of service overflow obtain information &  & \checkmark &  & \checkmark \\
14 & Directory traversal &  & \checkmark & \checkmark & \checkmark \\
15 & Execute code & \checkmark & \checkmark & \checkmark & \checkmark \\
16 & Execute code file inclusion & \checkmark &  &  &  \\
17 & Execute code gain privileges &  & \checkmark &  & \checkmark \\
18 & Execute code memory corruption &  & \checkmark & \checkmark & \checkmark \\
19 & Execute code memory corruption obtain information &  & \checkmark & \checkmark & \checkmark \\
20 & Execute code overflow & \checkmark & \checkmark & \checkmark & \checkmark \\
21 & Execute code overflow bypass a restriction or similar &  & \checkmark & \checkmark & \checkmark \\
22 & Execute code overflow memory corruption & \checkmark & \checkmark &  & \checkmark \\
23 & Execute code SQL injection & \checkmark &  &  &  \\
24 & Gain privileges &  & \checkmark &  & \checkmark \\
25 & Obtain information & \checkmark & \checkmark & \checkmark & \checkmark \\
26 & Overflow & \checkmark & \checkmark & \checkmark & \checkmark \\
27 & Overflow memory corruption & \checkmark & \checkmark &  & \checkmark \\
28 & Unspecified & \checkmark & \checkmark & \checkmark & \checkmark \\ \hline
 & \textbf{In total} & \textbf{18} & \textbf{24} & \textbf{16} & \textbf{25} \\ \hline
\end{tabular}%
}
\end{table}

\subsubsection{Bias in vulnerability distribution}
Vulnerability distribution refers to the numerical proportion of vulnerability types in a given dataset. Bias in the distribution could force a vulnerability prediction model to learn more from majority types during the training procedure, thus the model will perform poorly on minority types.

\begin{figure*}[htpb]
    \centering
    \subfigure[Asterisk]{
    \includegraphics[scale=0.5]{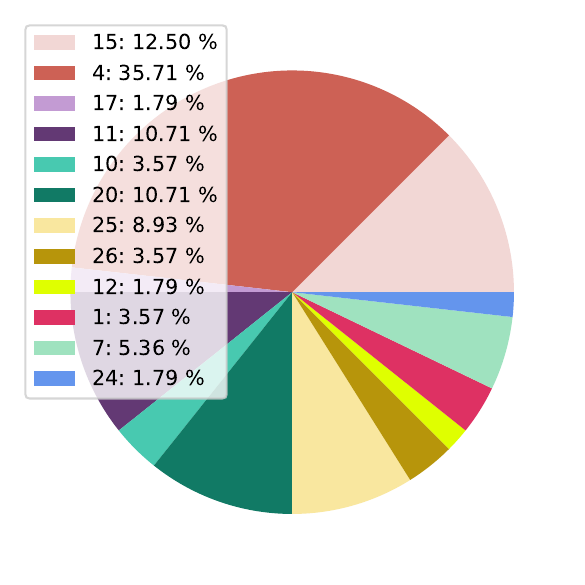}
    }
    \subfigure[FFmpeg]{
    \includegraphics[scale=0.5]{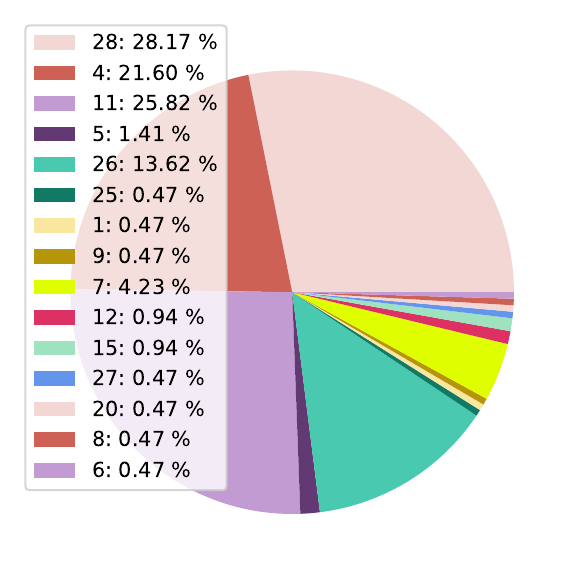}
    }
    \subfigure[LibPNG]{
    \includegraphics[scale=0.5]{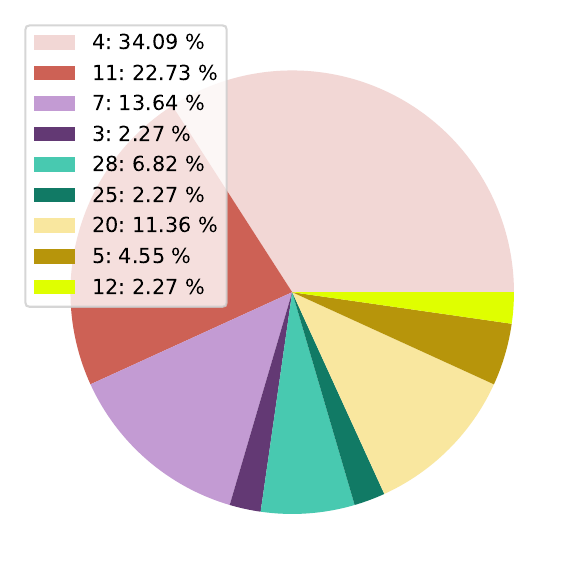}
    }
    \subfigure[LibTIFF]{\label{fig:vul_libtiff}
    \includegraphics[scale=0.5]{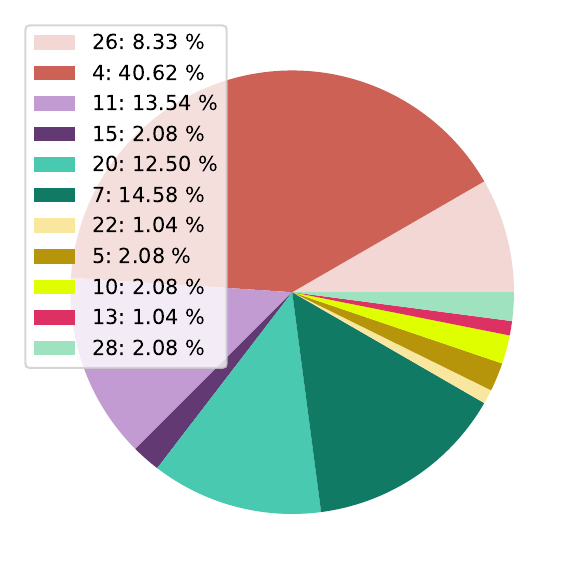}
    }
    \subfigure[Pidgin]{\label{fig:pidgin_libtiff}
    \includegraphics[scale=0.5]{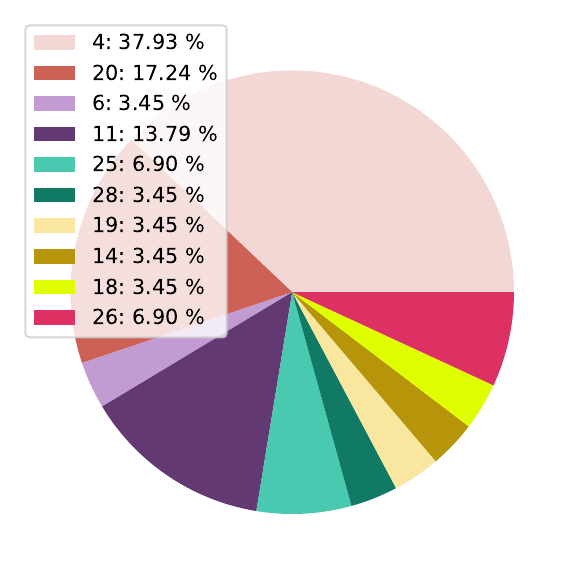}
    }
    \subfigure[VLC]{
    \includegraphics[scale=0.5]{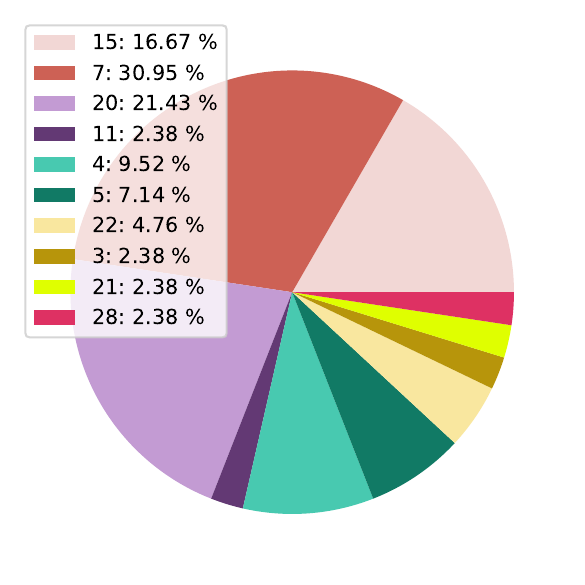}
    }
    \caption{Vulnerability distribution in each dataset provided by~\cite{lin2018cross}. Each color represents the vulnerability ID in Table~\ref{tab:vul_type} and the numeral text indicates the percentage of source code.}
    \label{fig:vul_distribution}
\end{figure*}

Figure~\ref{fig:vul_distribution} shows the vulnerability distribution of six datasets provided by~\cite{lin2018cross}. The first observation is that each dataset covers a different set of vulnerability types, confirming the defined coverage issue in Section~\ref{subsubsec:coverage}. Second, regardless of the dataset, vulnerabilities are distributed unevenly. Remarkably, in FFmpeg, nine out of 15 vulnerability types occupy less than 1\% of source code from the entire dataset. 

\subsection{Secondary issues}
This type of issue is mainly caused by errors in raw data or outdated information, which can be addressed through a comprehensive pre-processing to avoid affecting the model performance.

\subsubsection{Errors in raw data}
\label{subsec:errors}
Errors in raw data can significantly affect the accuracy of vulnerability detection algorithms. From existing datasets, we have found errors including empty source code files, extra lines, and inconsistent file formats. 

\cite{lin2018cross} includes 435, 195, and 205 empty source code files in Asterisk, Pidgin, and VLC, respectively. Without being informed, these empty files will be processed normally to train a detection model. However, since they do not provide any pattern, the detection model to be trained can be misled to learn real patterns for secure and vulnerable source code. 

All the 18 datasets provided by \cite{Lin2017discovery}, \cite{lin2018cross}, \cite{Lin2020deep}, and \cite{Lin2021multi} have the issue of containing an extra line at the beginning of a separate source code file. This extra line varies a lot across different datasets, such as \cite{Lin2021multi}: such as ``\verb|}|'', ``\verb|*/|'', ``\verb|} EightBpsContext;|'', ``\verb|};|'', ``\verb|} AascContext;|'', and ``\verb|)|''. Note that, without checking the source code files manually, this extra line is impossible to be removed during the pre-processing procedure. For example, the data pre-processing procedure generally filters comments by locating paired comment marks, such as ``\verb|//|'' and ``\verb|\* */|'' for C programming language. 

\cite{Lin2017discovery} uses '.txt' for vulnerable code files and '.c' for secure code files. Depending on the framework, reading files can encounter errors.

\subsubsection{Mislabeling on source code}
Mislabeling can severely degrade the performance of prediction models, which occurs due to the labeling manner. Generally, code files are extracted from open-source projects from GitHub and labels are manually given based on commit messages and descriptions in the national vulnerability database (NVD)~\cite{nvd_web}. Both commit messages and NVD entries are manually curated and analyzed, which is error-prone even with experienced software developers~\cite{lifelong2022}. Another way is to use static code analysis tools as shown in Table~\ref{tab:statis}, which is less accurate than the manual manner.

\begin{figure}
    \centering
    \includegraphics[scale=0.35]{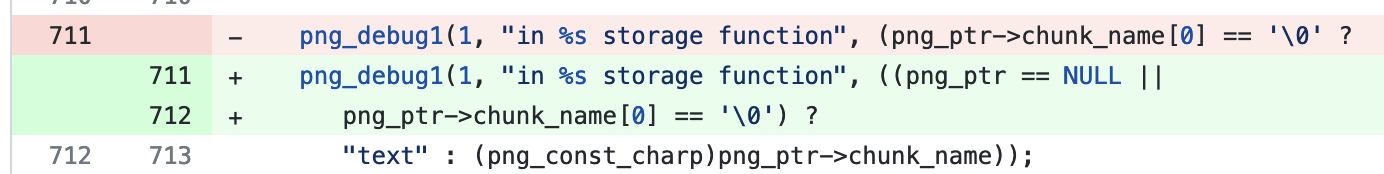}
    \caption{An example of mislabeling.}
    \label{fig:mislabeling}
\end{figure}

Figure~\ref{fig:mislabeling} shows an example of mislabeling in the LibPNG dataset provided by~\cite{lin2018cross}. The code file (including green lines in the example) named ``cve-2016-10087.c'' indicates that the source code is vulnerable that allows context-dependent attackers to cause a NULL pointer dereference vectors. However, according to the commit message on GitHub\footnote{More details can be found at: \url{https://github.com/glennrp/libpng/commit/a4d439b97507b54d7f08543e03eb8f006ea73bc5?diff=unified}}, the green lines are the patch of the corresponding vulnerable code (in red), thus, the label should be secure instead of vulnerable.

\subsubsection{Noisy historical data}
Noisy historical data~\cite{garg2022learning,Jimenez2019importance} refers to the phenomenon that code labeled as secure might be identified as vulnerable in the future given that most vulnerabilities are discovered much later than when they are introduced. For example, the \verb|decode_main_header()| function in FFmpeg is recently reported to have the null pointer dereference flaw\footnote{\url{https://nvd.nist.gov/vuln/detail/CVE-2022-3341}}. However, when collecting data before the report, this function will be considered as secure. 

\section{Good Practices}
\textbf{Using existing datasets:}
Critical issues are difficult to be addressed by data pre-processing, but there are techniques to mitigate their impact on the model performance. For the sampling size issue, data augmentation~\cite{mixcode2023} can be applied to increase the number of data during training. To force the model learn more from vulnerable code in imbalanced datasets, weighted loss functions, such as focal loss~\cite{focal2020lin} and mean squared error loss~\cite{seperated2016wang}, can be applied instead of the default cross entropy loss. Concerning the low vulnerability coverage issue, one can consider to merge several datasets that cover different vulnerabilities or just focus on detecting source code with included vulnerabilities. Last, code refactoring~\cite{yu2022data} and adversarial code attacks~\cite{alert2022} can help to generate more similar code samples without changing the semantics to reduce the bias in vulnerability distribution.

For secondary issues, the quality of existing datasets can be improved by designing an advanced pre-processing method to remove errors and reduce noise from the raw data. Note that, the errors mentioned in Section~\ref{subsec:errors} do not exist in all studied datasets and may not cover all cases. A thorough check should be made to develop the operations for a comprehensive pre-processing. To reduce noise, one should look into latest commit messages related to each code file and modify the labels when necessary.

\textbf{Creating new datasets:} When collecting source code from open-source projects, all found issues should be kept in mind. An exhaustive pre-processing step is highly recommended to avoid errors and noise in data. When assigning labels to collected data, both static analysis tools and expert manual manner should be considered to avoid mislabeling. In addition, one should include as much information as possible, such as vulnerability type, source project, and commit id, instead of only including labels to allow for tracking and re-checking.   

\section{Conclusion}
The accuracy and efficacy of deep learning (DL)-based vulnerability detection models highly rely on the quality of data used for training. Poor data quality can lead to unreliable results, false positives, and false negatives. In this paper, we define three critical and three secondary issues that occur in existing datasets. Furthermore, we provide actionable guidance to assist researchers in addressing these issues when using existing datasets or creating new ones with high quality.  

\section*{Acknowledgements}
This work was supported by the European Commission under the Horizon Europe Programme, as part of the project LAZARUS (\url{https://lazarus-he.eu/}) (Grant Agreement no. 101070303).

\bibliographystyle{plain}
\bibliography{refs}
\end{document}